\documentclass[conference]{IEEEtran} 
\ifCLASSOPTIONcompsoc
  \usepackage[nocompress]{cite}
\else
  \usepackage{cite}
\fi
\ifCLASSINFOpdf
\else
\fi
\usepackage[colorlinks=true,
            linkcolor=red,
            urlcolor=blue,
            citecolor=blue]{hyperref}
\usepackage{amsmath}
\usepackage{graphicx}
\usepackage{mathtools}
\usepackage{relsize}
\usepackage{graphicx}
\usepackage[scientific-notation=true]{siunitx}
\usepackage{algorithm}
\usepackage{filecontents,lipsum}
\usepackage{cite}
\usepackage{comment}
\usepackage{multirow}
\usepackage{amsfonts}
\usepackage[noend]{algpseudocode}
\usepackage[utf8]{inputenc}
\usepackage{graphicx}
\usepackage{caption}
\usepackage{latexsym}
\usepackage{etoolbox}
\usepackage{subcaption, float}
\makeatletter
\usepackage[inline]{enumitem}
\def\BState{\State\hskip-\ALG@thistlm}
\makeatother

\AfterEndEnvironment{table}{\vskip-1ex}

\newcommand{\qu}[1]{``#1''}

\DeclareMathOperator*{\argmax}{arg\,max}

\begin{document}

\title{CeliacNet: Celiac Disease Severity Diagnosis on Duodenal Histopathological Images Using Deep Residual Networks}

%
%

\markboth{IEEE Transactions on Medical Imaging}%
{Shell \MakeLowercase{\textit{et al.}}: Bare Demo of IEEEtran.cls for IEEE Journals}
%




\author{\IEEEauthorblockN{Rasoul Sali \IEEEauthorrefmark{1},
Lubaina Ehsan\IEEEauthorrefmark{2},
Kamran Kowsari\IEEEauthorrefmark{1}, 
Marium Khan\IEEEauthorrefmark{2},
Christopher A. Moskaluk\IEEEauthorrefmark{2},\\
Sana Syed\IEEEauthorrefmark{2}\IEEEauthorrefmark{3}, and
Donald E. Brown\IEEEauthorrefmark{1}\IEEEauthorrefmark{3}}

\IEEEauthorblockA{\IEEEauthorrefmark{1} Department of System and Information Engineering,
University of Virginia,
Charlottesville, VA, USA}

\IEEEauthorblockA{\IEEEauthorrefmark{2} Department of Pediatrics, School of Medicine, 
University of Virginia,
Charlottesville, VA, USA}

\IEEEauthorblockA{\IEEEauthorrefmark{3} School of Data Science, 
University of Virginia,
Charlottesville, VA, USA}

\{\href{mailto:rs8wa@virginia.edu}{rs8wa},
\href{mailto:lubaina@virginia.edu}{lubaina},
\href{mailto:kk7nc@virginia.edu}{kk7nc}, 
\href{mailto:mk2ne@virginia.edu}{mk2ne}, 
\href{mailto:cam5p@virginia.edu}{cam5p},
\href{mailto:sana.syed@virginia.edu}{sana.syed},
\href{mailto:deb@virginia.edu}{deb}\}@virginia.edu\vspace{-15pt}}

\maketitle

\begin{abstract}~Celiac Disease (CD) is a chronic autoimmune disease that affects the small intestine in genetically predisposed children and adults. Gluten exposure triggers an inflammatory cascade which leads to compromised intestinal barrier function. If this enteropathy is unrecognized, this can lead to anemia, decreased bone density, and, in longstanding cases, intestinal cancer.  The prevalence of the disorder is \textbf{$1\%$} in the United States. An intestinal (duodenal) biopsy is considered the “gold standard” for diagnosis. The mild CD might go unnoticed due to non-specific clinical symptoms or mild histologic features. In our current work, we trained a model based on deep residual networks to diagnose CD severity using  a histological scoring system called the modified Marsh score. The proposed model was evaluated using an independent set of \textbf{$120$} whole slide images from~\textbf{$15$}~CD patients and achieved an AUC greater than~$0.96$ in all classes. These results demonstrate the diagnostic power of the proposed model for CD severity classification using histological images.\vspace{5pt}
\end{abstract}


\section{Introduction}\label{sec:Introduction}~Celiac disease (CD) is an inability to normally process dietary gluten (present in foods such as wheat, rye, and barley) and is present in $1$\% of the US population. Gluten consumption by people with CD can cause diarrhea, abdominal pain, bloating, and weight loss. If unrecognized, it can lead to anemia, decreased bone density, and, in longstanding cases, intestinal cancer~\cite{fasano2003prevalence, parzanese2017celiac}. An intestinal (duodenal) biopsy, obtained via endoscopic evaluation, is considered the “gold standard” for diagnosis of CD.  Due to unclear clinical symptoms and/or obscure histopathological features~(based on biopsy images),~CD is often undiagnosed~\cite{corazza2007comparison}. There has been major clinical interest towards developing new and innovative methods to automate and enhance the detection of morphological features of CD on biopsy images.

Studies have shown the ease of training Convolutional Neural Networks (CNNs) for image recognition. These networks are a family of machine learning architectures which have proven to have superior performance over a wide range of computer vision tasks such as classification and object detection. Due to the wide availability of robust open source software and high-quality public datasets, these architectures are fast becoming the standard choice for being selected as the backbone of many modern computer vision technologies. Using large amounts of data, these models have shown to be effective in solving many biomedical imaging challenges. Currently, CNNs have been successfully applied to medical images such as MRI and X-rays~\cite{gulshan2016development, litjens2017survey}. CNNs have also shown promising performance on histopathological images~\cite{kowsari2019diagnosis,Mohammad_al_boni}.

Among various architectures of CNNs, Residual Networks~(ResNet) have received special attention due to their considerably superior performance in the analysis of histopathological images for disease detection, diagnosis and prognosis prediction to complement the opinion of a human pathologist. Multiple groups have published on the use of the ResNet architecture for classification of Hematoxylin and Eosin~(H\&E) stained biopsy images including breast and prostate cancer~\cite{gandomkar2018mudern,rakhlin2018deep,motlagh2018breast,chougrad2018deep,schaumberg2018h} and colorectal polyps~\cite{korbar2017deep}. Similarly impressive results for CD diagnosis based on whole slide biopsy images have been noted in published literature~\cite{wei2019automated}. Herein we explore the performance of deep residual networks in severity diagnosis of CD on duodenal biopsy images. 

This paper is organized as follows: In Section~\ref{sec:severitytypes}, disease severity classes of CD are presented. In Section~\ref{sec:Data_Source}, we describe the  data used in this study. Section~\ref{sec:Pre-Processing} presents the data pre-processing steps. The methodology is explained in Section~\ref{sec:Method}. Empirical results are elaborated in Section~\ref{sec:Empirical_Results}. Finally, Section~\ref{sec:Conclusion} concludes the paper along with outlining future directions.

\section{Severity Classes of Celiac Disease}\label{sec:severitytypes}
\begin{figure}[!t]
    \centering
    \includegraphics[width=0.95\columnwidth]{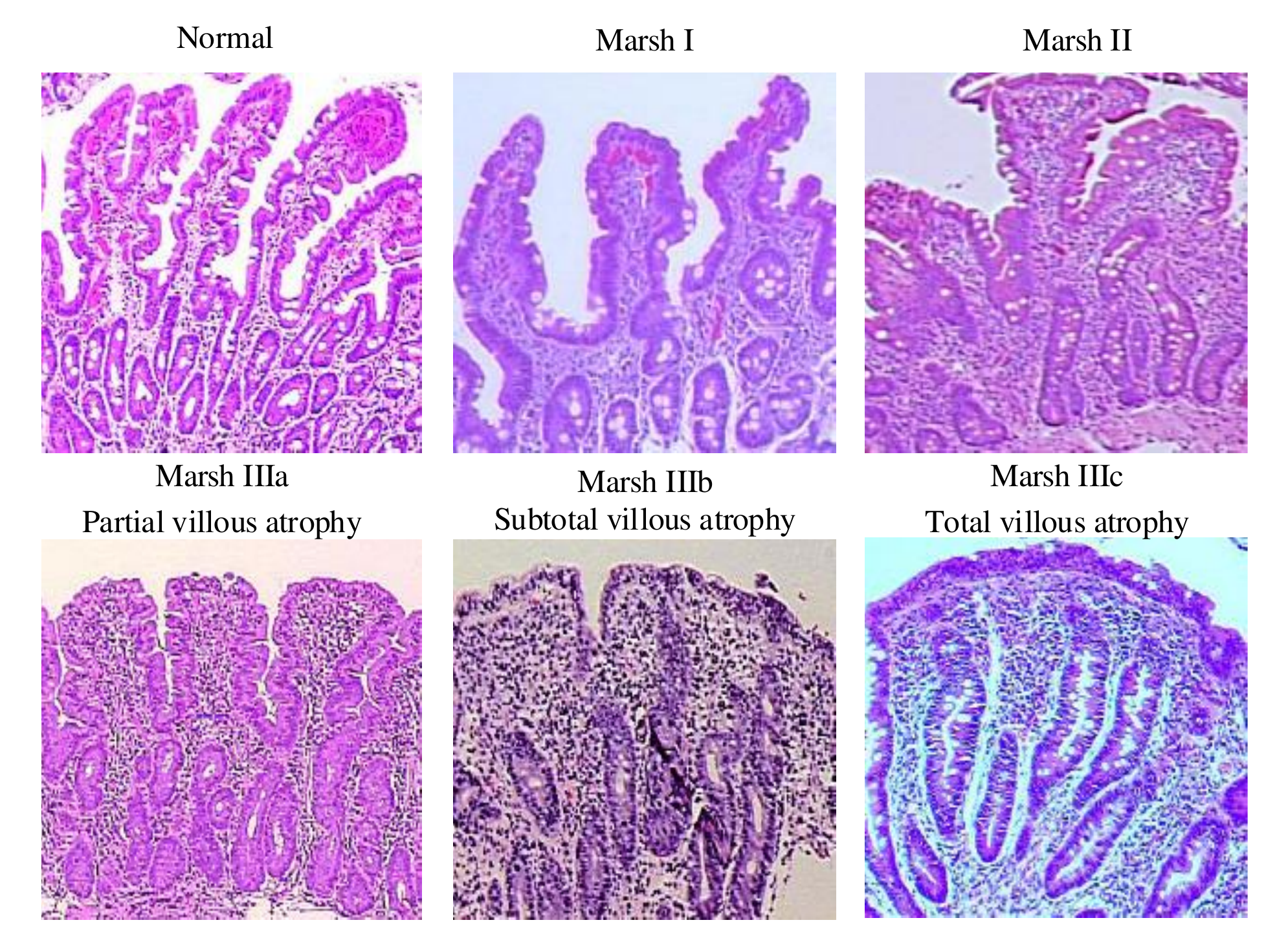}
    \vspace{-5pt}
    \caption{CD severity classification based on modified Marsh score~\cite{fasano2001current}} \label{fig:MarshScore}
    \vspace{-15pt}
\end{figure}

~Modified Marsh Score Classification was developed to classify the severity of CD based on microscopic histological morpological features~(Figure~\ref{fig:MarshScore}). It takes into account the architecture of the duodenum as having finger-like projections~(called \qu{villi}) which are lined by cells called epithelial cells. Between the villi are crevices called crypts that contain regenerating epithelial cells. 
The normal ratio of the length of a typical healthy villus to the depth of a representative health crypt  should be between 3:1 and 5:1. In the normal, healthy duodenum (first part of the small intestine), there should be no more than 30 immune cells known as lymphocytes interspersed per 100 epithelial cells in the top layer of the villus. Marsh~I histology comprises of normal villus architecture with an increase in the number of intraepithelial lymphocytes. Marsh~II includes increased intraepithelial lymphocytes along with a finding known as crypt hypertrophy in which the crypts appear enlarged. This is usually rare since patients typically rapidly progress from Marsh I to IIIa. Marsh~III is sub-divided into~IIIa~(partial villus atrophy), Marsh IIIb~(subtotal villus atrophy) and Marsh IIIc~(total villus atrophy) to explain the spectrum of villus atrophy along with crypt hypertrophy and increased intra-epithelial lymphocytes. Finally, in Marsh~IV, villi are completely atrophied. This is called “hypoplastic” or complete villus atrophy and describes the microscopic histology of duodenal tissue from patients at the extreme end of gluten sensitivity.

\section{Data Source}\label{sec:Data_Source} 
~$162$ H\&E stained duodenal biopsy slides were obtained from the archival biopsies of $34$ CD patients from the University of Virginia~(UVa) in Charlottesville, VA, United States. Each slide contained multiple biopsies per patient resulting in $336$ whole slide images at 40x magnification using the Leica SCN 400 slide scanner~(Meyer Instruments, Houston, TX) at the Biorepository and Tissue Research Facility at UVa. Characteristics of our patient population were as follows: the median $(Q1, Q3)$ age was $130~(92.5, 175.5)$ months. we had a roughly equal  distribution  of males  $(47.1\% ,n=16)$  and  females  $(52.9\%, n=18)$. Biopsy images for our study population were scored by two medical professionals and validated with reads from a pathologist specialized in gastroenterology. Our biopsy image dataset ranged from Marsh I to IIIc with no biopsy images present in Marsh II.

\section{Data Pre-processing}\label{sec:Pre-Processing}~ Since whole slide images (WSIs) were digitized at high resolutions, these were large files with notable color variability apparent on visual inspection. Therefore, we pre-processed these before any computational analyses were conducted. This section describes all pre-processing steps including image patching, patch clustering and color normalization. 
\subsection{Image patching}\label{subsec:Patching}~The effectiveness of CNNs in image classification has been shown in various studies across different domains~\cite{kowsari2018rmdl,hu2018deep,heidarysafa2018improvement}. However, the training of a CNN on high resolution WSIs that are at a gigapixel level is not often feasible due to high computational cost. Also, the application of CNNs on WSIs further contributes to the loss of a large part of discriminatory information due to extensive down-sampling which is needed in such images~\cite{hou2016patch}. We hypothesized that since there were cellular level morphological differences between different CD severity classes given the spectrum of pathology, a trained classifier on image patches would likely perform as well or better than a trained WSI-level classifier.
A sliding window method was applied to each high-resolution WSI to generate patches of size $500\times500$ pixels with~$50\%$ overlapping area. After generating patches from each image, we labelled each patch based on its associated image.

\begin{figure}[!b]
    \centering
    \vspace{-16pt}
    \includegraphics[width=0.95\columnwidth]{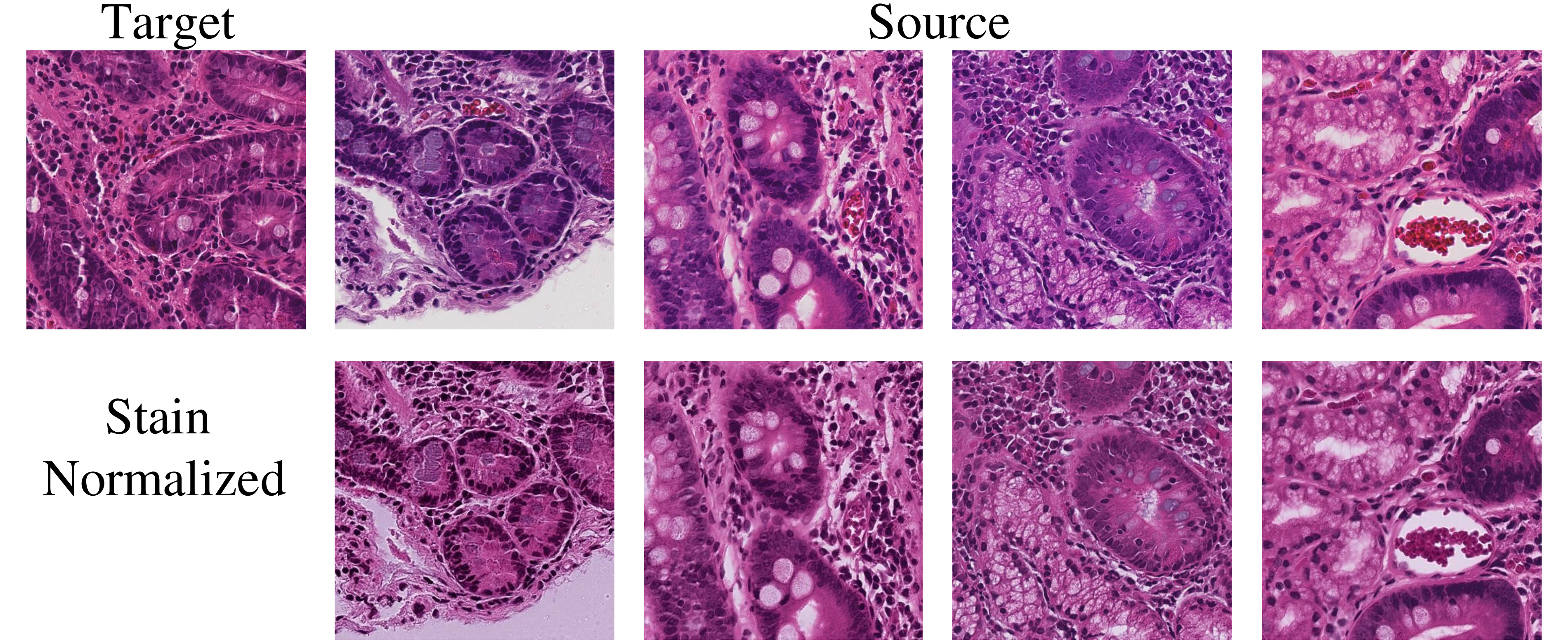}
    \caption{Color normalization artifacts when using the method proposed by Vahadane et al.~\cite{vahadane2016structure}. Images in the first row represent the target image and some source images. Their associated normalized images are in second row} \label{fig:StainNormalization}
\end{figure}

\subsection{Patch Clustering}\label{subsec:Clustering}~Clustering is organizing objects in a such way that objects within a group or cluster in some way are more similar to each other compared to objects in other groups. There is a wide variety of algorithms for data clustering and K-means clustering is one of the easiest ones~\cite{jain2010data}. Finding the optimal solution to the k-means clustering problem is NP-hard in general Euclidean space even for 2 clusters. Clustering of $n$ $d$-dimension entities in $k$ clusters can be exactly solved in time of $O(n^{dk+1})$~\cite{aloise2009np}. Obviously, reduction of dimension $d$ will result in significant improvement of the K-means clustering algorithm in term of time complexity. To address the problem of dimensionality reduction, a convolutional auto-encoder~\cite{goodfellow2016deep} was used to learn embedded features of each patch. These auto-encoders have been reported in the literature as having had great success as a dimensionality reduction method via the powerful reprehensibility of neural networks~\cite{wang2014generalized}.

\begin{figure}[!t]
    \centering
    \includegraphics[width=0.98\columnwidth]{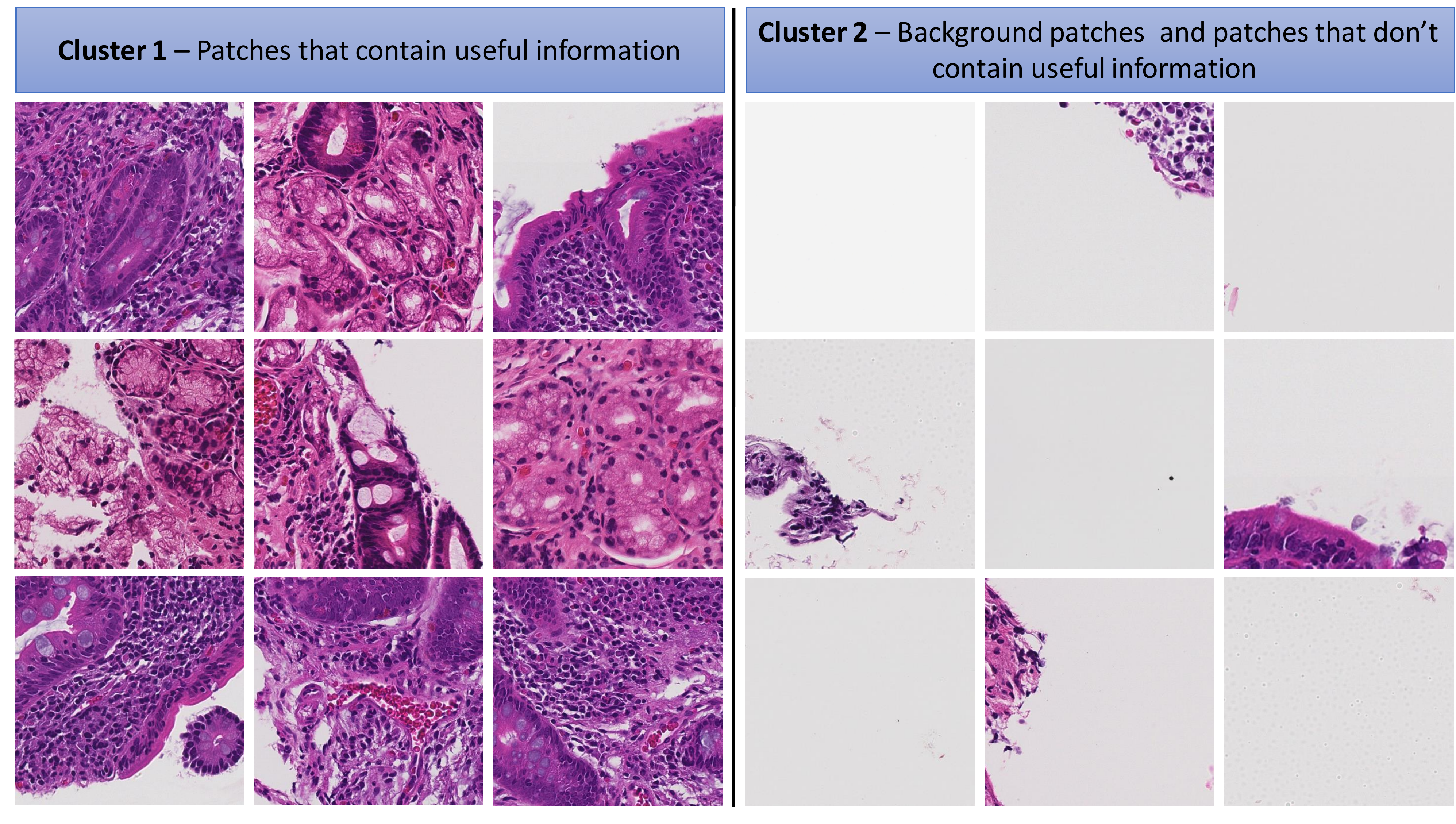}
    \caption{Some samples of clustering results - cluster 1 included patches with useful information and cluster 2 included patches without useful information~(mostly created from the slide background and border areas of the WSIs)} \label{fig:clusters}
\end{figure}

In our work, a two-step clustering process was applied to identify useless patches which had mostly been created from the background of the WSIs. All or a large part of these patches were blank or did not  contain any useful biopsy information. Through the first step, a convolutional autoencoder was used to learn the embedded features of each patch and in the second step k-means clustering algorithm was applied to cluster embedded features into two clusters: useful and not useful. Some results of patch clustering have been shown in Figure~\ref{fig:clusters}.

\subsection{Stain normalization}\label{subsec:StainNormalization}~Histological images have substantial color variation that adds bias while training the model. This arises due to a wide variety of factors such as differences in raw materials and manufacturing techniques of stain vendors, staining protocols of labs, and color
responses of digital scanners~\cite{vahadane2016structure}. To avoid any bias, unwanted color variations are neutralized by conducting color normalization as an essential pre-processing step prior to any analyses. 

Various color normalization approaches have been proposed in the published literature. In this study, we used the approach proposed by Vahadane et al.~\cite{vahadane2016structure}. This approach preserves
biological structure information by basing color mixture modeling on sparse non-negative matrix factorization. Figure~\ref{fig:StainNormalization} shows an example of the result of applying this  technique on representative biopsy patches.

\section{Methodology}\label{sec:Method}
\subsection{Model development}\label{subsec:Model}
\begin{figure}[!t]
    \centering
    \includegraphics[scale=0.3]{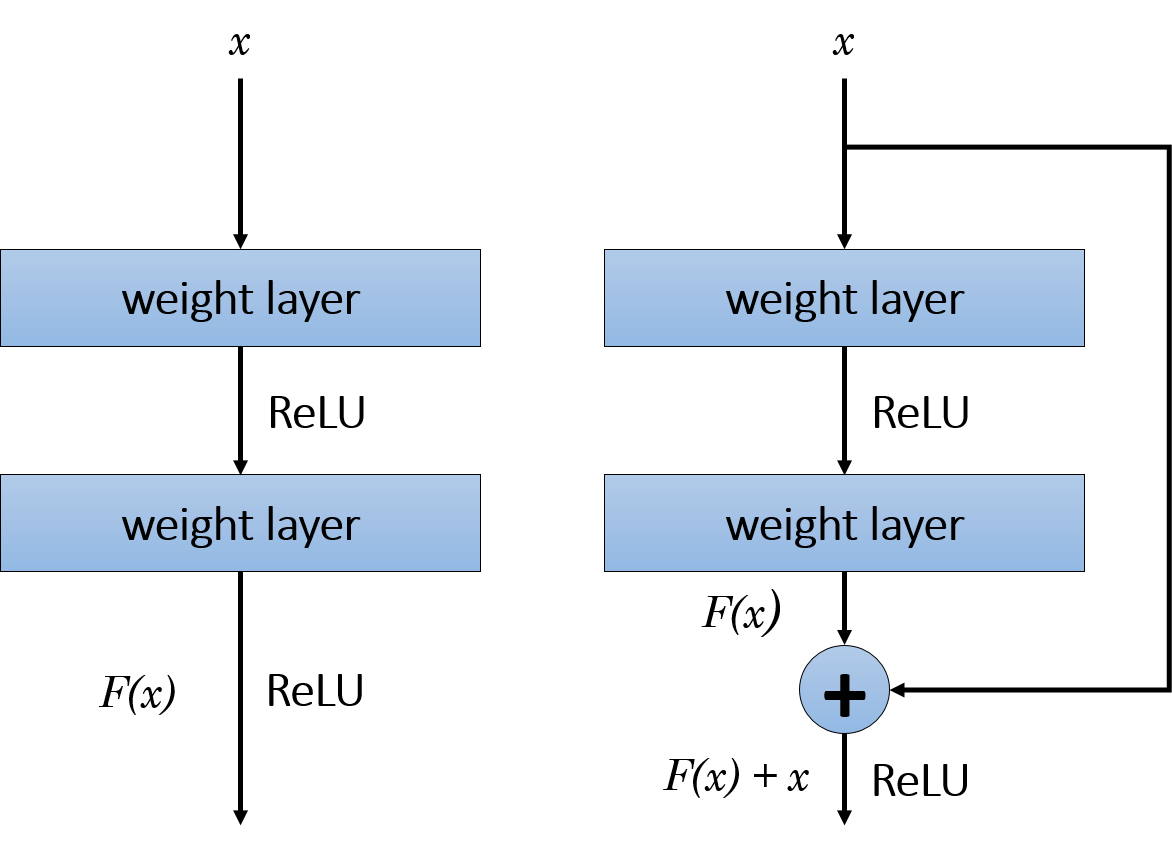}
    \caption{Building blocks of~(left) a traditional CNN,~(right) a ResNet } \label{fig:buildin blocks}
\end{figure}

~ CNNs have demonstrated promising performance in image classification tasks. There are many different architectures of CNNs in the literature,with associated advantages and drawbacks. In the current study, we used the deep residual network~(ResNet)\cite{he2016deep}, a model which has shown great performance in image classification problems including medical image analysis\cite{gandomkar2018mudern,wen2018deep}. Although it has been shown that CNNs with more convolutional layers achieve the most accurate results, simply stacking more convolutional layers will not lead to better performance. When the deep network reaches a certain depth, its performance tends to be saturated and even begins to rapidly decline. In such cases, the models involve a large amount of parameters and are computationally expensive to train through whole parameters. This is called the degradation problem and ResNet was originally proposed to tackle this issue. The core idea of ResNet is introducing a skip connection that skips one or more layers and bypasses the input from the previous layer to the next layer without any modification. Since these added shortcut connections perform identity mapping, extra parameters are not added to the model. Such architecture enables deployment of deeper networks without problem of degeneracy. The building block of the ResNet is compared to the building block of the traditional network in Figure \ref{fig:buildin blocks}. In the traditional networks, the mapping from input to output can be represented by the nonlinear function $H(x)$. In residual learning blocks, $F(x) = H(x) - x$ is used as mapping function~\cite{he2016deep}. In essence, as part of traditional CNNs, the input $x$ is mapped to $F(x)$ which is a completely new representation that does not keep any information about the original input, while ResNet blocks compute a slight change to the original input $x$ to get a slightly altered representation. ResNet was the Winner of the~ILSVRC~2015 in image classification, detection, and localization, as well as the Winner of the MS COCO~2015 detection and segmentation.

Different variants of ResNet models such as ResNet50, ResNet101, and ResNet152 were trained on the ImageNet dataset~\cite{guillaumin2012large}. We customized the Resnet50 by removing fully connected layers and keeping only the ResNet backbone as a feature extractor. Then we added one fully connected layer with $1024$ neurons that received the flattened output of the feature extractor. Finally, the output layer was added such that it represented a  prediction  probability  for  each  of  the  four Marsh score categories: I, IIIa, IIIb and IIIc. We used dropout on the fully-connected layers with $p=0.5$ as the regulizer. This model has been summarized in Table~\ref{tb:model}.

\begin{table}[H]
\caption{Architecture of the model}\label{tb:model}
\centering
\begin{tabular}{cccc}
\hline
\textbf{Class} & \textbf{Layer Type}  & \textbf{Output Shape}  & \textbf{Number of Parameters}  \\ \hline

 1 & Model& $(7,7,2048)$  & $2,3587,712$ \\
 
 2 & Flatten & $100352$  & $0$ \\
 
 3 & Dense & $1024$  & $102,761,472$ \\
 
 4 & Dropout & $1024$  & $0$ \\
 
 5 & Dense & $4$  & $4,100$ \\\hline
\end{tabular}
\end{table}

\begin{table*}[!t]
\centering
\caption{Patch-level performance of model for celiac disease severity diagnosis}\label{table:patch-level results}
\noindent\begin{tabular}{ccccc}
\hline 
\textbf{~~~~~~~Class~~~~~~~} & \textbf{~~~~~~~Accuracy~(\%)~~~~~~~}  & \textbf{~~~~~~~Precision~(\%)~~~~~~~}  & \textbf{~~~~~~~Recall~(\%)~~~~~~~}  & ~~~~~~~\textbf{F1-measure~(\%)}~~~~~~~ \\ \hline

 I$~(n = 6988)$ & $89.54~(88.82,~90.26)$& $93.30~(92.71,~93.89)$  & $89.54~(88.82,~90.26)$ & $91.38~(90.72,~92.04)$  \\
 
 IIIa$~(n = 6615)$ & $84.75~(83.88,~85.61)$ & $94.16~(93.59,~94.73)$  & $84.75~(83.88,~85.62)$ & $89.20~(88.45,~89.95)$  \\
 
 IIIb$~(n = 7695)$ & $89.45~(88.76,~90.13)$ & $83.94~(83.12,~84.76)$  & $89.45~(88.76,~90.14)$ & $86.61~(85.85,~87.37)$  \\
 
 IIIc$~(n = 7369)$ & $90.61~(89.94,~91.28)$ & $85.53~(84.73,~86.33)$  & $90.61~(89.94,~91.28)$ & $87.99~(87.25,~88.73)$ \\\hline 
\end{tabular}
\vspace{-5pt}
\end{table*}

We resized pre-processed patches into~$224\times224$ pixels and used them to train our model. Both horizontal and vertical random rotations were performed as part of our data augmentation. The model was trained on around~$50,000$ patches for each of four classes. Optimization was performed using RMSprop optimization with
no momentum, a base learning rate of \num{1e-5} and a multiclass cross entropy loss function.

\begin{figure}[!b]
    \centering
    \vspace{-10pt}
    \includegraphics[width=\columnwidth]{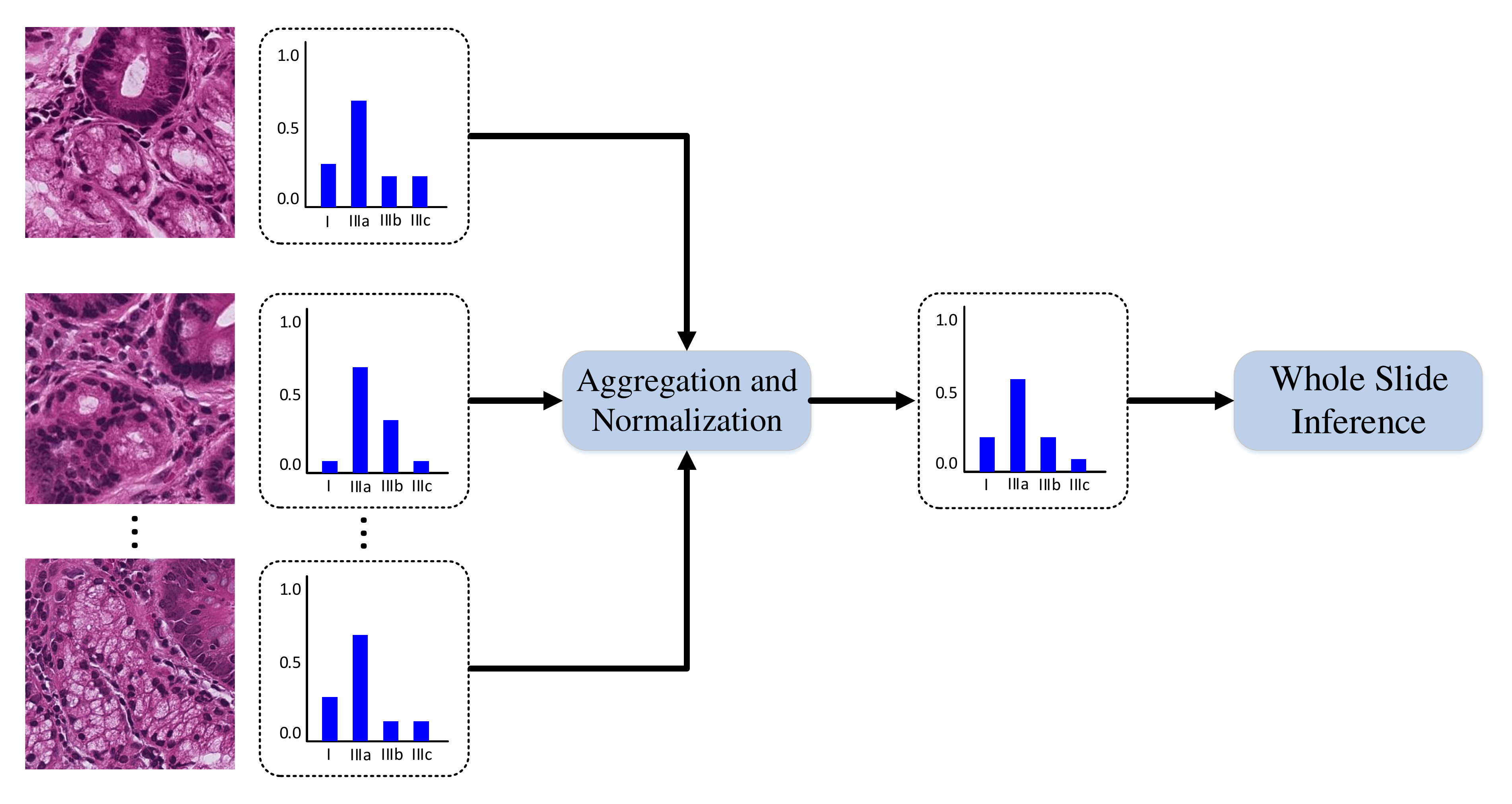}
    \caption{Overview of whole-slide inference process using aggregation of patch-level classifications} \label{fig:WSIClassification}
\end{figure}

\subsection{Whole slide classification}\label{sec:WSIClassification}~Our goal was to classify WSIs based on severity assessed via the modified Marsh score. The model used was trained to classify small patches rather than WSIs. To achieve this goal, a heuristic method was developed which aggregated crop classifications and translated them to whole-slide inferences. Each WSI in the test set was initially patched, those patches which did not contain any information were filtered out and finally stain normalization was performed. After these pre-processing steps our trained model was applied with the goal of image classification. We denoted the probability distribution over possible labels, given the crop $x$ and training set~$D$ by~$p(y\vert x,D)$. In general, this represented a vector of length $C$, where $C$ is number of classes. In our notation, the probability is conditional on the test patch $x$, as well as the training set~$D$. For each crop, the model gives an output of a vector composed of four components showing probabilities for each one of the four classes of CD severity. Given a probabilistic output, the patch~$j$ in slide $i$ is assigned to the most probable class label~$\hat{y}_{ij}$ which is shown in Equation~\ref{eq:patchClass}.

\begin{equation}\label{eq:patchClass}
\hat{y}_{ij} = \argmax_{c \in \{1,2,3,...,C\}} p(y_{ij} = c\vert x_{ij},D)
\end{equation}

where $\hat{y}$ is called maximum a posteriori~(MAP) . Summation over these vectors and normalizing the resultant vector, created a vector that had components showing the probability of CD severity for the associated WSI. Equation~\ref{eq:slideClass}, shows how the class of WSI was predicted. 

\begin{equation}\label{eq:slideClass}
\hat{y}_i = \argmax_{c \in \{1,2,3,...,C\}} \sum_{j=1}^{N_i}p(y_{ij} = c\vert x_{ij},D)
\end{equation}

where $N_i$ is number of patches in slide $i$. Figure~\ref{fig:WSIClassification} depicts overview of the whole-slide inference process.

\begin{figure}[!b]
\vspace{-20pt}
    \centering
    \includegraphics[width=0.9\columnwidth]{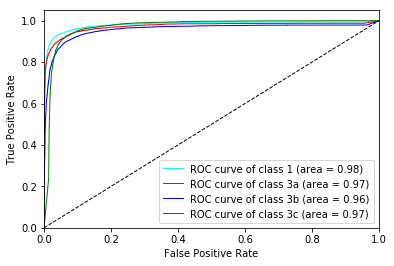}
    \caption{Patch-level ROC and AUC for different classes} \label{fig:ROC}
\end{figure}

\begin{figure*}[!htb]
    \centering
    \includegraphics[width=0.99\textwidth]{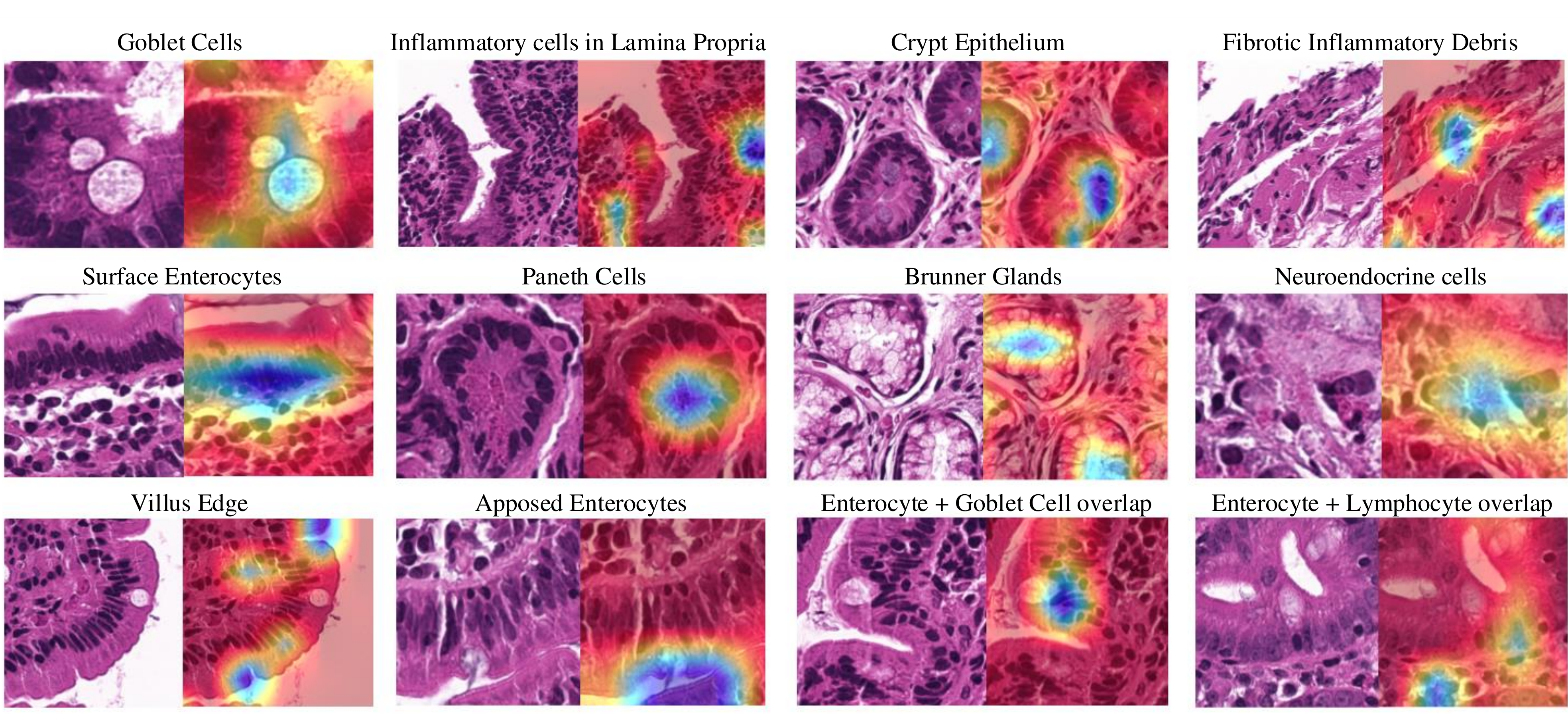}
    \caption{ Class activation mapping heat maps highlighting the most informative regions of patches relevant to different categories including goblet cells, inflammatory cells in lamina propria, crypt epithelium, fibrotic inflammatory debris, surface enterocytes, Paneth cells, Brunner's glands, neuroendocrine cells, villus edge and apposed enterocytes. Area of attention is shown in blue color.
} \label{fig:gradcam_results}
\vspace{-5pt}
\end{figure*}

\section{Experimental results}\label{sec:Empirical_Results}
\subsection{Patch-level performance}~To evaluate the effectiveness of our proposed model, we used an independent test set including~$120$ WSIs. After application of a sliding window for patching these whole slides and doing the aforementioned pre-processing steps,~$28,667$ crops remained to be used for our model  evaluation. Performance of our model on this set is shown
in Table \ref{table:patch-level results}, which includes accuracy, precision, recall, and
the F1 score with~$95\%$ confidence intervals. Also patch-level ROC curves and AUC for each class are shown in Figure~\ref{fig:ROC}. As shown AUC for all classes was greater than $0.96$.

\subsection{Slide-level performance}
After classification of the test patches, their results were aggregated based on the method described in section~\ref{sec:WSIClassification} to make an inference about each test slide. By applying this method, all slides in the test set were classified correctly and the accuracy of the model in all the classes was $100\%$. In the four classes of I, IIIa, IIIb and IIIc there were $20$, $21$, $44$ and $35$ slides, respectively. This means that CD severity was correctly diagnosed.

\subsection{Class Activation Mapping}~We used the Grad-CAM approach to obtain visual explanation microscopic feature heat-maps for WSI patch areas predictive of CD severity. Grad-CAM visualizations were obtained for~$350$ images~($95$ Marsh I, $75$ Marsh IIIa,~$100$ Marsh~IIIb,~$80$ Marsh IIIc). Qualitatively, the Grad-CAM images of our model localized microscopic morphological features such as different cell types and tissue structures that corresponded to the disease pathology. Quantitatively, our Grad-CAM heat-maps were reviewed by two medical professionals. These heat-maps were broadly categorized into~10 groups that are as follows:  goblet cells, inflammatory cells in the lamina propria, crypt epithelium, fibrotic inflammatory debris, surface enterocytes, Paneth cells, Brunner's glands, neuroendocrine cells, villus edge and apposed enterocytes. Visualization of these different categories on individual patches are shown in Fig~\ref{fig:gradcam_results}. Most images depicted an overlap of heat-map for enterocytes and goblet cells or enterocytes and lymphocytes that are known to be representative of CD~\cite{oberhuber1999histopathology}~(Fig~\ref{fig:gradcam_results}).

\subsection{Hardware and Framework}
All of the results shown in this paper are performed on Central Process Units~(CPU) and Graphical Process Units~(GPU). Also, This model is capable to be performed on only GPU, CPU, or both.The processing units that has been used through this experiment was intel on \textit{Xeon~E5-2640~ (2.6 GHz)} with~\textit{12 cores} and \textit{64~GB} memory~(DDR3). Also, graphical card on our machine is \textit{Nvidia Quadro~K620} and \textit{Nvidia Tesla~K20c}. This work is implemented in Python using Compute Unified Device Architecture~(CUDA) which is a parallel computing platform and Application Programming Interface ~(API) model created by $Nvidia$. We used $TensorFelow$ and $Keras$ library for creating the neural networks~\cite{abadi2016tensorflow,chollet2015keras}.

\section{Conclusion}\label{sec:Conclusion}~In this paper, we investigated CD severity using CNNs applied to histopathological images. A state-of-the-art deep residual neural network architecture was used to categorize patients based on H\&E stained duodenal histopathological images into four classes, representing different CD severity based on a histological classification called the modified Marsh score. Our model was trained to classify different patches of WSIs. In addition we provided a heuristic to aggregate results of patch classification and make inference about the WSIs. Our model was tested on~$28,667$ crops derived from an independent test set $120$ WSIs from $15$ CD patients. It achieved AUC greater than $0.96$ in all classes. At the WSI level classification, the proposed model correctly classified all WSIs. Validation results were highly promising and showed that our model has great potential to be utilized by pathologists to support their CD severity decision based on a histological assessment. We  also used the  Grad-CAM approach to obtain  visual explanation of microscopic features predictive of CD severity. These heat-maps were broadly categorized into~$10$ groups including goblet cells, inflammatory cells in the lamina propria, crypt epithelium, fibrotic inflammatory debris, surface enterocytes, Paneth cells, Brunner's glands, neuroendocrine cells, villus edge and apposed enterocytes.

Albeit achieving promising results, this study has a number of limitations. Firstly, healthy cases were not included this study. This is an avenue for future work. In addition, all biopsy images used in this study were collected from a single medical center and scanned with the same equipment, thus our
data may not be representative of the entire range of histopathologic patterns in patients worldwide. Furthermore, the target image for stain normalization was selected manually based on the opinion of a pathologist. Selecting a different image as the target image could affect the appearance of stain normalized images. It is known that some variability exists in this selection, which is then propagated through the framework. Finally, in this study we applied a single method of stain normalization and the use of other methods may lead to different results. Therefore, investigating the effect of different stain normalization techniques can be another potential area of future work. 

\section*{Acknowledgements}

This research was initially supported by an Engineering in Medicine SEED Grant from the University of Virginia~$(SS~\&~DEB)$ and the University of Virginia Translational Health Research Institute of Virginia~($THRIV$) Mentored Career Development Award~$(SS)$. Research reported in this publication was supported by [National Institute of Diabetes and Digestive and Kidney Diseases] of the National Institutes of Health under award number~$[K23~DK117061-01A1]$. The content is solely the responsibility of the authors and does not necessarily represent the official views of the National Institutes of Health.

\bibliographystyle{IEEEtran} 
\bibliography{refs}

\end{document}